# Analytic representation of discrete and continuous mechanical systems

Benoy Talukdar [1], Supriya Chatterjee [2] and Sekh Golam Ali [3]

[1] Department of Physics, Visva-Bharati University, Santiniketan 731235, India

[2] Department of Physics, Bidhannagar College, EB-2, Sector-1, Salt Lake, Kolkata 700064, India

Department of Physics, Kazi Nazrul University, Asansol 713304, India

**Abstract**

We investigate how the theory of self-adjoint differential equations alone can be used to provide a satisfactory solution of the inverse vatiational problem. For the discrete system, the self-adjoint form of the Newtonian equation allows one to find an explicitly time-dependent Lagrangian representation. On the other hand, the same Newtonian equation in conjunction with its adjoint forms a natural basis to construct an explicitly time-independent analytic representation of the system. This approach when applied to the equation of damped harmonic oscillator help one disclose the mathematical origin of the Bateman image equation. We have made use of a continuum analog of the same approach to find the Lagrangian or analytic representation of nonlinear evolution equations.

## 1. Introduction

In point mechanics the term 'analytic representation' refers to description of Newtonian systems by means of Lagrangians [1]. Understandably, to find the analytic representation of a mechanical system one begins with the equation of motion and then constructs a Lagrangian function by using a strict mathematical procedure discovered by Helmholtz [2]. In the calculus of variation this is the so-called inverse problem which is more complicated than the usual direct problem where one first assigns a Lagrangian function using phenomenological consideration and then computes the equation of motion using the Euler-Lagrange equation [3]. However, there are two types of analytic representation, namely, the direct and indirect ones. We can introduce the basic concepts of direct and indirect analytic representations by using a system of two uncoupled harmonic oscillators with equations of motion

$$\ddot{q}(t) + \omega^2 q(t) = 0 \tag{1}$$

and
$$\ddot{y}(t) + \omega^2 y(t) = 0. \tag{2}$$

It is straightforward to verify that the system of equations (1) and (2) can be analytically represented either by the Lagrangian

$$L_d = \frac{1}{2}\left(\dot{q}^2(t) + \dot{y}^2(t)\right) - \frac{\omega^2}{2}\left(q^2(t) + y^2(t)\right) \tag{3}$$

or by the Lagrangian

$$L_i = \dot{q}(t)\dot{y}(t) - \omega^2 q(t) y(t). \tag{4}$$

Here overdots denote differentiation with respect to time $t$. The function $L_d$ refers to the Lagangian giving the direct analytic representation of the system presumably because it yields the equation of motion for $q(t)$ ($y(t)$) via the Euler-Lagrange equation written in terms of $q(t)$ ($y(t)$). On the other hand, $L_i$ yields the equation of motion for $q(t)$ ($y(t)$) via the Euler-Lagrange equation written in terms of $y(t)$ ($q(t)$). This is why the representation of the system by the use of $L_i$ is called indirect analytic representation. This simple example indicates that Lagrangian representations of Newtonian systems are not unique. The problem of non-uniqueness of the Lagrangian functions has deep consequences for the correspondence between symmetries and constants of the motion. For example, the direct Lagrangian (3) is rotationally invariant such that the associated Noether constant of the motion is the angular momentum. As opposed to this the indirect Lagrangian (4) is invariant under 'squeeze' transformation $(q(t), y(t) \to (q(t)e^t, y(t)e^{-t})$. Consequently, for the Lagrangian $L_i$, conservation of angular momentum is associated with the invariance under squeeze [4]. Moreover, the Hamiltonians corresponding to direct and indirect Lagrangians provide two distinct routes for quantization of open systems [5-12].

In this work we shall first present a general method to construct direct and indirect analytic representations of Newtonian systems by using the theory of self-adjoint differential equations [13]. In particular, we shall demonstrate that the self-adjoint form of a Newtonian equation can always be used to find its direct analytic representation even if the original equation does not satisfy the Helmholtz criteria [2]. On the other hand, any Newtonian equation in conjunction with its adjoint allows one to construct its indirect analytical representation. We shall then turn our attention to similar problems in classical field theory with particular attention to Lagrangian representations of nonlinear evolution equations which play a role in soliton theory [14]. In the recent past Ibragimov [15] constructed indirect analytic representations for a number of such equations by introducing the concept of nonlinirar self-adjointness. In this context we note that almost simultaneously with the work of Ibragimov, two of us [16] derived an elegant method to write adjoint equations for nonlinear evolution equations having at least one conserved density and thus constructed their indirect analytic representation. Since all physically important nonlinear evolution equations are characterized by a number of conserved densities, the mathematical framework used by us is also quite general. Therefore, it appears that the problem of finding indirect analytic representation of nonlinear evolution equations has been solved quite satisfactorily. In view of this, we shall be interested here to look for direct analytic representation of physically interesting nonlinear evolution equations.

A single nonlinear evolution equation is never an Euler-Lagrange expression [17] so as to follow from a Lagangian (possibly direct) involving the field variable $u(x,t)$ and its space and time derivatives. One common trick to convert such an equation into the variational form is to replace $u(x,t)$ by a potential function (often called the Casimir potential) defined by

$$\omega(x,t) = \int_x^\infty dy\, u(y,t), \tag{5}$$

For instance, in view of (5) it is rather straightforward to recast the Korteweg-de Vries (KdV) equation

$$u_t = 6uu_x + u_{3x} \tag{6}$$

in the variational form [14]

$$\delta \int_{t_1}^{t_2} dt \int_{-\infty}^{\infty} dx\, \Im(\omega_t, \omega_x, \omega_{2x}) = 0 \tag{7}$$

so as to define a Lagrangian density

$$\Im = \frac{1}{2}\omega_t \omega_x + \frac{1}{2}\omega_{2x}^2 + \omega_x^3 \tag{8}$$

for the KdV equation. Each equation in the KdV hierarchy constitutes a Lagrangian system. However, there are many nonlinear evolution equations which do not admit direct analytic representation [18]. We shall first focus our attention on the first few members of the KdV hierarchy and provide a method to compute results for their Lagrangian densities for them , and subsequently adapt the approach to deal with nonlinear evolution equations which do not allow straightforward analytic representation.

In sec. 2 we demonstrate how the theory of self-adjoint differential equations can be judiciously used to provide a complete solution of the inverse variational problem in Newtonian mechanics. It is well known that dissipative forces cannot be accommodated within the framework of variational principle. This is true even for the simple-minded damped harmonic oscillator. We have shown that the self-adjoint form of the damped-oscillator equation automatically leads to direct analytic representation. However, the associated Lagrangian is explicitly time dependent. Additionally, the equation of the damped harmonic oscillator in conjunction with its adjoint can be used to construct an indirect analytic representation which does not depend explicitly on time. We then provide in Tables 1and 2 the results for direct and indirect analytic representations for a number of ordinary differential equations which play an important role in physical theories. In sec.3 we adapt the approach followed in sec.2 to deal with problems of continuum mechanics. In particular, we construct the direct Lagrangian representation for a number of physically interesting nonlinear evolution equations in (1+1) dimension. Finally, in sec.4 we summarize our outlook of the present and make some concluding remarks.

## 2. Analytic representation of Newtonian systems

From the inverse problems in the calculus of variations [1] one knows that all Newtonian systems cannot have Lagrangian representation. In particular, the equations written in the general form

$$F_i = A_{ij}(t,q,\dot{q})\ddot{q}^j + B_i(t,q,\dot{q}) = 0, \qquad i,j = 1,2,....n \text{ and } q = q(t) \subset R^n. \tag{9}$$

will have a Lagrangian representation if and only if

$$\frac{\partial F_i}{\partial \ddot{q}^j} = \frac{\partial F_j}{\partial \ddot{q}^i}, \tag{10a}$$

$$\frac{\partial F_i}{\partial \dot{q}^j} + \frac{\partial F_j}{\partial \dot{q}^i} = 2\frac{d}{dt}\left(\frac{\partial F_i}{\partial \ddot{q}^j}\right) \tag{10b}$$

and 
$$\frac{\partial F_i}{\partial q^j} - \frac{\partial F_j}{\partial q^i} = \frac{1}{2}\frac{d}{dt}\left(\frac{\partial F_i}{\partial \dot{q}^j} - \frac{\partial F_j}{\partial \dot{q}_l}\right). \tag{10c}$$

The relations (10a) – (10c) are often called the Helmholtz conditions [2] and give the necessary and sufficient conditions for the existence of a Lagrangian function for any Newtonian system. Equation (9) represents an N-dimensional differential equation. For the one-dimensional case (10a) and (10c) become identity such that in this case we are left with only one condition

$$\frac{\partial F}{\partial \dot{q}} = \frac{d}{dt}\left(\frac{\partial F}{\partial \ddot{q}}\right) \qquad (11)$$

for the Lagrangian representation of a one-dimensional system represented by

$$F = p(t)\ddot{q}(t) + r(t)\dot{q}(t) + s(t)q(t) = 0 \ . \qquad (12)$$

Equation (12) will satisfy the Helmholtz condition (11) if $r(t) = \dot{p}(t)$. The equation of motion then becomes

$$\frac{d}{dt}\left(p(t)\dot{q}(t)\right) + s(t)q(t) = 0 \ . \qquad (13)$$

Multiplying (13) by $\delta q$ and integrating over $t$ from $t_1$ to $t_2$ we can recast it in the variational form

$$\delta \int_{t_1}^{t_2}\left(\frac{p(t)}{2}\dot{q}^2(t) - \frac{s(t)}{2}q^2(t)\right)dt = 0 \qquad (14)$$

such that (12) follows from the Lagrangian function

$$L = \frac{p(t)}{2}\dot{q}^2(t) - \frac{s(t)}{2}q^2(t) \qquad (15)$$

via the Euler-Lagrange equation

$$\frac{d}{dt}\left(\frac{\partial L}{\partial \dot{q}(t)}\right) - \frac{\partial L}{\partial q(t)} = 0. \qquad (16)$$

For arbitrary values of $p(t)$ and $r(t)$, (12) is not self-adjoint. However, the theory of linear second-order, self-adjoint differential equations is quite general [13]. For example, we can always transform (12) in the self-adjoint form $F_{sadj}$ by multiplying it with a non-vanishing factor

$$\rho(x) = \frac{1}{p(t)} e^{\int (r/p)dt} \qquad (17)$$

such that

$$F_{sadj} = \frac{d}{dt}\left(\dot{q}(t) e^{\int \frac{r(t)}{p(t)}dt}\right) + \frac{s(t)q(t)}{p(t)} e^{\int \frac{r(t)}{p(t)}dt} = 0. \qquad (18)$$

On the other hand, the adjoint equation $F_{adj}$ corresponding to (11) can be found by changing the independent variable by

$$q(t) = y(t) e^{-\int ((r(t)-\dot{p}(t))/p(t))dt} . \tag{19}$$

We thus have

$$F_{adj} = a(t)\ddot{y}(t) + b(t)\dot{y}(t) + c(t)y(t) = 0 \tag{20}$$

with

$$a(t) = p(t), \quad b(t) = 2\dot{p}(t) - r(t) \quad \text{and} \quad c(t) = \ddot{p}(t) - \dot{r}(t) + s(t) . \tag{21}$$

Understandably, the equation will be self-adjoint if $F = F_{adj}$. It is of interest to note that (i) the self-adjoint equation (18) can be judiciously used to provide a time-dependent Lagrangian representation of (12) even when (12) violates the Helmholtz condition. On the other hand, (12) in conjunction with its adjoint (20) provides a basis (ii) to write indirect Lagrangian for (12).

For simplicity of presentation we shall first demonstrate (i) and (ii) by taking recourse to the use of the damped Harmonic oscillator with equation of motion given by

$$F = m\ddot{q}(t) + \gamma \dot{q}(t) + kq(t) = 0, \tag{22}$$

where $m$ and $k$ stand for the mass and spring constant of the oscillator and the symbol $\gamma$ represents the frictional coefficient of the medium in which the oscillation takes place. From (11) and (21) $p(t) = m, r(t) = \gamma$ and $s(t) = k$. Thus from (18) we can write the self-adjoint equation for (22) in the form

$$F_{sadj} = e^{\frac{\gamma t}{m}} \left( \ddot{q}(t) + \frac{\gamma}{m} \dot{q}(t) + \frac{k}{m} q(t) \right) = 0 . \tag{23}$$

In close analogy with the derivation of (14) it is straightforward to recast (23) in the Hamilton's variational form with the Lagrangian given by

$$L = \frac{e^{\frac{\gamma t}{m}}}{m} \left( \frac{1}{2} m \dot{q}^2(t) - \frac{1}{2} k q^2(t) \right). \tag{24}$$

About 75 years ago, the Hamiltonian corresponding to the Lagrangian in (24) was used independently by Caldirola [5] and by Kanai [6] to quantize the damped harmonic oscillator. In the recent past it has been shown that [19] there are some typical problem in their quantization procedure presumably because the oscillator in (22) represents an open system that continuously gives energy to the surrounding. There is another representation of dissipative system which consists in treating the damped oscillator together with an amplified oscillator such that the energy drained out from the first is completely absorbed by the second. As early as 1931 this model was introduced by Bateman [20]. Understandably, we have now a dual system which is closed. Bateman dual system has also been used to study quantum dissipation with some added advantage [9, 19] over the model of Caldirola and of Kanai. The amplified oscillator associated with the damped system (22) is written as

$$F' = m\ddot{y}(t) - \gamma \dot{y}(t) + ky(t) = 0. \tag{25}$$

Mechanistically, this attempt to understand dissipation by the simultaneous use of (22) and (25) amounts to doubling the degrees of freedom to study the problem. Using $F$ and $F'$ we write a Lagrangian given by

$$L_i = \left( y(t)F + q(t)F' - m\frac{d}{dt}(y(t)\dot{q}(t) + q(t)\dot{y}(t)) \right)/2 \tag{26}$$

which on simplification reads

$$L_i = m\dot{q}(t)\dot{y}(t) + \frac{\gamma}{2}(q(t)\dot{y}(t) - \dot{q}(t)y(t)) - kq(t)y(y). \tag{27}$$

In writing (26) we have assumed that Lagrangian of a system is its own equation of motion [4]. The third term in (26) is a trivial Gauge term [21] of a second-order Lagrangian and has been introduced only to write a first-order Lagrangian for the system. It is easy to see that (27) provides an indirect analytic representation of the damped harmonic oscillator. The image equation (25) of the damped harmonic oscillator was introduced by Bateman [20] using purely phenomenological arguments. It is an interesting mathematical curiosity to note that (25) is the adjoint equation of (22). This can easily be proved by making use of (20) and (21). Thus we see that given a second-order differential equation and its adjoint one can always construct an indirect analytical representation of the system.

From the self-adjoint forms of (22) and (25) it is also possible to find a direct Lagrangian

$$L_d = \frac{e^{\frac{\gamma t}{m}}}{m}\left( \frac{1}{2}m\dot{q}^2(t) - \frac{1}{2}kq^2(t) \right) + \frac{e^{-\frac{\gamma t}{m}}}{m}\left( \frac{1}{2}m\dot{y}^2(t) - \frac{1}{2}ky^2(t) \right) \tag{28}$$

for the Bateman dual system. Similarly, we can write the results for direct and indirect Lagrangians for the coupled quartic anharmonic or Duffing oscillators

$$\ddot{x}_1(t) = -\omega^2 x_1(t) - 4\alpha x_1^3(t) - 12\alpha x_1(t)x_2^2(t) \tag{29}$$

and
$$\ddot{x}_2(t) = -\omega^2 x_2(t) - 4\alpha x_2^3(t) - 12\alpha x_2(t)x_1^2(t) \tag{30}$$

in the form

$$L_d = \frac{1}{2}\left(\dot{x}_1^2(t) + \dot{x}_2^2(t)\right) - \frac{1}{2}\omega^2\left(x_1^2(t) + x_2^2(t)\right) - \alpha\left(x_1^4(t) + x_2^4(t)\right) - 6\alpha\left(x_1^2(t)x_2^2(t)\right) \tag{31}$$

and $\quad L_i = \dot{x}_1(t)\dot{x}_2(t) - \omega^2 x_1(t)x_2(t) - 4\alpha\left(x_1^3(t)x_2(t) + x_2^3(t)x_1(t)\right) \tag{32}$

respectively.

The methods outlined above for the damped harmonic oscillator can easily be extended to derive direct and indirect analytical representations of other second-order linear differential equations which have important applications in physical theories. In Tables 1 and 2 we present results for Lagrangians giving direct and indirect analytic representations of seven such equations. In presenting the results we always use $t$ as the independent variable and, for brevity, call it as time. Column 2 in Table1 gives the self-adjoint equation corresponding to the original equation in column 1.

Table1. Self-adjoint differential equations and Lagrangians giving direct analytic representayion of some important linear second-order differential equation of mathematical Physics.

| Original differential equation | Self-adjoint equation | Lagrangian giving direct analytic representation |
|---|---|---|
| Legendre $(1-t^2)\ddot{q}(t) - 2t\dot{q}(t) + n(n+1)q(t) = 0$ | $(1-t^2)\ddot{q}(t) - 2t\dot{q}(t) + n(n+1)q(t) = 0$ | $\frac{1}{2}(1-t^2)\dot{q}^2(t) - \frac{1}{2}n(n+1)q^2(t)$ |
| Bessel $t^2\ddot{q}(t) + t\dot{q}(t) + (t^2 - v^2)q(t) = 0$ | $t\ddot{q}(t) + \dot{q}(t) + \left(t - \frac{v^2}{t}\right)q(t) = 0$ | $\frac{1}{2}t\dot{q}^2(t) - \frac{1}{2}\left(t - \frac{v^2}{t}\right)q^2(t)$ |
| Laguerre $t\ddot{q}(t) + (1-t)\dot{q}(t) + nq(t) = 0$ | $e^{-t}(t\ddot{q}(t) + (1-t)\dot{q}(t) + nq(t)) =$ | $\frac{1}{2}te^{-t}\dot{q}^2(t) - \frac{1}{2}ne^{-t}q^2(t)$ |
| Hermite $\ddot{q}(t) - 2t\dot{q}(t) + 2nq(t) = 0$ | $e^{-t^2}(\ddot{q}(t) - 2t\dot{q}(t) + 2nq(t)) = 0$ | $\frac{1}{2}e^{-t^2}\dot{q}^2(t) - ne^{-t^2}q^2(t)$ |
| Chebyshev $(1-t^2)\ddot{q}(t) - t\dot{q}(t) + n^2q(t) = 0$ | $\dfrac{-((1-t^2)\ddot{q}(t) - t\dot{q}(t) + n^2q(t))}{\sqrt{t^2-1}} = 0$ | $\frac{1}{2}\sqrt{t^2-1}\dot{q}^2(t) + \frac{1}{2}n^2q^2(t)/\sqrt{t^2-1}$ |
| Gaussian Hypergeomertic $(1-t)t\ddot{q}(t) + ((1+\alpha+\beta)t - \gamma)\dot{q}(t) + \alpha\beta q(t) = 0$ | $(1-t)^{-2-\alpha-\beta+\gamma}t^{-1-\gamma}((1-t)t\ddot{q}(t) + ((1+\alpha+\beta)t-\gamma)-\gamma)\dot{q}(t) - \alpha\beta q(t)) = 0$ | $t^{-c}(1-t)^{-1+\alpha+\beta+\gamma}\left(\dfrac{\frac{1}{2}\dot{q}^2(t) +}{\dfrac{\alpha\beta q^2(t)}{2t(1-t)}}\right)$ |
| Confluent Hypergeometric $t\ddot{q}(t) + (\gamma - t)\dot{q}(t) - \alpha q(t) = 0$ | $e^{-t}t^{c-1}(t\ddot{q}(t) - (\gamma-t)\dot{q}(t) - \alpha q(t)) = 0$ | $t^c e^{-t}\left(\frac{1}{2}\dot{q}^2(t) + \dfrac{\alpha q^2(t)}{2t}\right)$ |
| | | |

The results for the direct Lagrangians are presented in column 3. The Legendre equation is self-adjoint such that the equations in columns 1 and 2 of row 1 are same. On the other hand, the other equations in the Table are non-self-adjoint. Consequently, for these equations the self-adjoint forms are different from the parent equations. All Lagrangians giving the direct analytic representation are explicitly time dependent and closely resemble the result in (24) for the damped harmonic oscillator. Hurestically, one can construct expressions for the Lagrangians in Column 3 from the corresponding self-adjoint equations by following a recipe that reads

"Neglect the term involving $\dot{q}(t)$ in the self-adjoint form of the differential equation and use the mapping $\ddot{q}(t) \to \frac{1}{2}\dot{q}^2(t)$ and $q(t) \to -\frac{1}{2}q^2(t)$ in the first and third terms"

In close analogy with the results displayed in Table 1 we reserve columns 1 and 2 of Table2 for the original equations and their adjoints. In column 3 we present results for Lagrangians giving indirect analytic representation for the same set of equations as considered in Table 1. Looking closely into the entries of Table 2 we see that the Legendre equation and its adjoint are same. This result is quite expected since Legendre equation is self-adjoint. The Lagrangian function for the Legendre equation is of the same form as that in (27) for the damped harmonic oscillator except that the Lagrangian does not involve any term analogous to the middle term in (27). This is, however, not true for other equations in the Table, which are not self-adjoint. For example, the Lagrangian functions for all other equations have middle terms in the form $\beta(t) = y(t)\dot{q}(t) - q(t)\dot{y}(t)$. Interestingly, the indirect Lagrangians of all self-adjoint differential is free from the $\beta(t)$ and appear in the form of the Lagrangian function for the Legendre equation.

Table2. Adjoint differential equations and Lagrangians giving indirect analytic representation of the linear second-order differential equations in Table 1

| Original differential equation | Adjoint equation | Lagrangoan giving indirect analytic reptesentation |
|---|---|---|
| Legendre $(1-t^2)\ddot{q}(t) - 2t\dot{q}(t) + n(n+1)q(t) = 0$ | $(1-t^2)\ddot{y}(t) - 2t\dot{y}(t) + n(n+1)y(t) = 0$ | $(1-t^2)\dot{q}(t)\dot{y}(t) - n(n+1)q(t)y(t)$ |
| Bessel $t^2\ddot{q}(t) + t\dot{q}(t) + (t^2-v^2)q(t) =$ | $t^2\ddot{y}(t) + 3t\dot{y}(t) + (1+t^2-v^2)y(t) = 0$ | $t^2\dot{q}(t)\dot{y}(t) + \frac{1}{2}t(y(t)\dot{q}(t) - q(t)\dot{y}(t)) - (t^2 - v^2 + \frac{1}{2})q(t)y(t)$ |
| Laguerre $t\ddot{q}(t) + (1-t)\dot{q}(t) + nq(t) = 0$ | $t\ddot{y}(t) + (1+t)\dot{y}(t) + (n+1)y(t) = 0$ | $t\dot{q}(t)\dot{y}(t) + \frac{1}{2}t(y(t)\dot{q}(t) - q(t)\dot{y}(t)) \left(n + \frac{1}{2}\right)q(t)y(t)$ |
| Hermite $\ddot{q}(t) - 2t\dot{q}(t) + 2nq(t) = 0$ | $\ddot{y}(t) + 2t\dot{y}(t) + 2(n+1)y(t) = 0$ | $\dot{q}(t)\dot{y}(t) + t(y(t)\dot{q}(t) - q(t)\dot{y}(t)) - (2n+1)q(t)y(t)$ |
| Chebyshev $(1-t^2)\ddot{q}(t) - t\dot{q}(t) + n^2 q(t) = 0$ | $(1-t^2)\ddot{y}(t) - 3t\dot{y}(t) - (1-n^2)y(t) = 0$ | $(1-t^2)\dot{q}(t)\dot{y}(t) + \frac{1}{2}t(q(t(\dot{y}(t) - y(t)\dot{q}(t))) + \left(\frac{1}{2} - n^2\right)q(t)y(t)$ |
| Gaussian Hypergeomertic $(1-t)t\ddot{q}(t) + ((1+\alpha+\beta)t - \gamma)\dot{q}(t) - \alpha\beta q(t) = 0$ | $(1-t)t\ddot{y}(t) + (2 - t(5+\alpha+\beta) + \gamma)\dot{y}(t) - (3+\alpha+\beta+\alpha\beta)y(t) = 0$ | $(1-t)t\dot{q}(t)\dot{y}(t) + \frac{1}{2}(1 - 3t - \alpha t - \beta t + \gamma) \times (\dot{q}(t)y(t) - \dot{y}(t)q(t)) + \frac{1}{2}(3+\alpha+\beta+\alpha\beta)q(t)y(t)$ |
| Confluent Hypergeometric $t\ddot{q}(t) + (\gamma - t)\dot{q}(t) - \alpha q(t) = 0$ | $t\ddot{y}(t) + (2 + t - \gamma)\dot{y} + (1-\alpha)y(t) = 0$ | $t\dot{q}(t)\dot{y}(t) + \frac{1}{2}(1+t-\gamma) \times ((y(t)\dot{q}(t) - \dot{y}(t)q(t) - \left(\frac{1}{2} - \alpha\right)q(t)y(t)$ |

## 3. Analytic representation of field equations

In classical mechanics we deal with systems with finite degrees of freedom such that the equations of motion follow from the action principle $\delta\int_{t_1}^{t_2} L(q_i(t), \dot{q}_i(t), t)dt$. Here the discrete index $i = 1,2,.....n$ and $L(.)$ stands for the Lagrangian of the system. In hydrodynamics, electrodynamics or in the theory of gravitation we have systems which possess an infinite number of degrees of freedom. The physical situation is then described by means of a field $u(x,t)$, $x \subset \Re^d, t \subset \Re$. A field theory is a generalization of classical mechanics in which the field variable $u(x,t)$ plays the role of the dynamic variable $q_i(t)$. The discrete index $i$, $1 \leq i \leq n$ now becomes the continuous variable $x \subset \Re^d$ and $\sum_{i=1}^{n}$ is replaced by $\int_{\Re^d} dx$. The action functional for (d+1) dimensional systems is written as $I = \int_{\Re^{d+1}} \pounds$, the variation of which leads to the appropriate Euler-Lagrange equations. Specializing to the (1+1) dimensional case we write the action principle in the form

$$\delta\int \pounds dxdt = 0 \tag{33}$$

so as to get The Euler-Lagrange equation

$$\frac{d}{dt}\left(\frac{\partial \Im}{\partial u_t}\right) - \frac{\delta \Im}{\delta u} = 0. \tag{34}$$

We are interested here in the analytic representation of nonlinear evolution equations which are of first order in time but higher order in the space variable. Keeping this in view we express the variational derivative in (34) as

$$\frac{\delta}{\delta u} = \sum_{k=0}^{n}(-1)^k \frac{\partial^k}{\partial u^k}\frac{\partial}{\partial u_{kn}}, \quad u_{kn} = \frac{\partial^k u}{\partial u^k}. \tag{35}$$

In (34) and (35) the subscripts on $u$ denote appropriate partial derivatives.

In continuum mechanics involving nonlinear operators the Helmholtz formulation of the inverse problem proceeds by considering an $r$-tuple of differential functions written as

$$P[u] = P(x, u^{(n)}) \in A^r \tag{36}$$

and then defining the so-called Fre'chet derivative [17]. The Fre'chet derivative of $P$ is the differential operator $D_P : A^q \to A^r$ so that

$$D_P(Q) = \frac{d}{d\varepsilon}\bigg|_{\varepsilon=0} P[u + \varepsilon Q[u]] \tag{37}$$

for any $Q \in A^q$. If

$$\wp = \sum_J P_J[u] D_J \tag{38}$$

is a differential operator, its adjoint is the differential operator $\wp^*$ given by

$$\wp^* = \sum_J (-D)_J \cdot P_j \tag{39}$$

meaning that for any $Q \in A$

$$\wp^* Q = \sum_J (-D)_J [P_J Q]. \tag{40}$$

Equation (40) provides a mere restatement of the transformation (17I in the frame of functional analysis. For example, identifying (38) with the differential operator of the damped harmonic oscillator (22) one can make use of (40) to obtain the Bateman's image equation (25). However, a differential operator $\wp$ is selfadjoint if $\wp^* = \wp$. The Helmholtz condition asserts that $P$ is the Euler-Lagrange expression for some variational problem iff $D_P$ is selfadjoint. When self-adjointness is guaranteed, a Lagrangian density for $P$ can be obtained explicitly using the homotopy formula

$$\Im[v] = \int_0^1 u P[\lambda v] d\lambda. \tag{41}$$

Equation (41) has been used in ref.18 to obtain direct analytic representation of nonlinear evolution equations. In the following we shall follow a different approach to find similar analytic representation.

More than a decade ago Desaix et al. [22] reported an accurate approximation solution of the Thomas-Fermi nonlinear differential equation

$$\frac{d^2\varphi(\chi)}{d\chi^2} = \frac{\varphi^{\frac{3}{2}}(\chi)}{\sqrt{\chi}} \tag{42}$$

by using a Lagrangian function given by

$$L = \frac{1}{2}\left(\frac{d\varphi(\chi)}{d\chi}\right)^2 + \frac{2}{5}\frac{\varphi^{\frac{5}{2}}(\chi)}{\sqrt{\varphi}}. \tag{43}$$

Here $\varphi(\chi)$ stands for the universal Thomas-Fermi function. The Lagrangian function in (43) can easily be obtained by expressing (42) in the variational form. But let us follow a different route to construct the Lagrangian function .We assume that the Lagrangian can be found from the equation of motion and thus write

$$L = \varphi(\chi)\left(\frac{d^2\varphi(\chi)}{d\chi^2} - \frac{\varphi^{\frac{3}{2}}(\chi)}{\sqrt{\varphi}}\right) - \frac{d}{d\chi}\left(\varphi(\chi)\frac{d\varphi(\chi)}{d\chi}\right). \tag{44}$$

It is easy to verify that (44) is not the correct Lagrangian for the Thomas-Fermi equation. In view of this we postulate that the linear and nonlinear terms in the equation occur in the Lagrangian with unequal weights such that the Lagrangian is now written as

$$L = \varphi(\chi)\left(a\frac{d^2\varphi(\chi)}{d\chi^2} - b\frac{\varphi^{\frac{3}{2}}(\chi)}{\sqrt{\chi}}\right) - \frac{d}{d\chi}\left(a\varphi(\chi)\frac{d\varphi(\chi)}{d\chi}\right). \tag{45}$$

The weight factors $a$ and $b$ can now be determined by demanding that the Lagrangian (45) when used in the Euler-Lagrange equation reproduce the associated equation of motion. Thus we arrive at the result for the Lagrangian used by Desiax et al. We shall use this approach to find direct analytic representation of nonlinear evolution equations. It may appear that we are using some kind of bootstrapping to achieve our goal. However, in the course of our study we shall see that the present model has some advantages over direct use of the homotopy formula (41). In any case the first step in looking for the Lagrangian representation of a given nonlinear evolution is to make sure that, written in terms of Casimir potential ,it is self adjoint.

In terms of the Casimir potential, the KdV equation (6) reads

$$\omega_{xt} = \omega_{4x} - 6\omega_x\omega_{2x}. \tag{46}$$

From (45) we write the Euler-Lagrange expression

$$P[\omega] = \omega_{4x} - 6\omega_x\sigma_{2x} \tag{47}$$

which in conjunction with (37) yields the Fre'chet derivative

$$D_P = D_x^4 - 6\omega_x D_x^2 - 6\omega_{2x} D_x \ldots \quad (48)$$

Using (39) it is easy to prove that $D_P = D_P^*$. The KdV equation thus involves a nonlinear self adjoint differential operator to follow from a Lagrangian. We now integrate (46) with respect to $x$ and make use of appropriate boundary conditions to write

$$\omega_t - \omega_{3x} + 3\omega_x^2 = 0. \quad (49)$$

In close analogy with (45) we now write the Lagrangian density for the KdV equation as

$$\Im = \omega_x(a(\omega_t - \omega_{3x}) + 3b\omega_x^2) + \frac{\partial}{\partial x}(a\omega_x\omega_{2x}) \quad (50)$$

and demand that it should yield (49) via an appropriate Euler-Lagrange equation similar to that given in (34).. This gives $a = \frac{1}{2}$ and $b = \frac{1}{3}$ and the Lagrangian density in (8).

Lax [23] discovered a family of nonlinear partial differential equations, each member of which is of higher order than that in (6) but shares some common properties with the KdV equation. These equations form the so-called KdV hierarchy. The second member in the hierarchy is given by

$$u_t - u_{5x} + 10 u u_{3x} + 20 u_x u_{2x} - + 30 u^2 u_x = 0. \quad (51)$$

Writing (51) in terms of Casimir potential and integrating over $x$ we get

$$\omega_t - \omega_{5x} - 10\omega_x\omega_{3x} - 5\omega_{2x}^2 - 10\omega_x^3 = 0 \quad . \quad (52)$$

As before we write the Lagrangian density as

$$\Im = \omega_x(a(\omega_t - \omega_{5x}) - 10b\omega_x\omega_{3x} - 5c\omega_{2x}^2 - 10d\omega_x^3) + \frac{\partial}{\partial x}(a\omega_x\omega_{4x}) \quad (53)$$

and demand that (53) via the Euler-Lagrange equation should give (52). We thus get

$$a = \frac{1}{2},\ d = \frac{1}{4} \text{ and } 4b - c = 1. \quad (54)$$

Equation (54) shows that the values of $a$ and $d$ are unique while the value of $b$ (or $c$) depends on the choice of $c$ (or $b$). The arbitrary choice of $b$ or $c$ leads to only gauge equivalent Lagrangians. For example, the values of the parameters $c = 0$ and $b = \frac{1}{4}$ give

$$\Im_1 = \frac{1}{2}(\omega_t\omega_x + \omega_{2x}\omega_{4x}) - \frac{5}{2}\omega_x^2\omega_{3x} - \frac{5}{2}\omega_x^4 \quad (55)$$

and
$$\Im_2 = \frac{1}{2}(\omega_t \omega_x + \omega_{2x}\omega_{4x}) - 5\omega_x^2 \omega_{3x} - 5\omega_x \omega_{2x}^2 - \frac{5}{2}\omega_x^4. \tag{56}$$

It is easy to check that $\Im_1$ and $\Im_2$ differ by a gauge term $\dfrac{\partial}{\partial x}\left(\dfrac{5}{2}\omega_x^2 \omega_{2x}\right)$. The method described here can easily be adapted to find analytic representation of other members in the KdV hierarchy

The KdV and higher KdV equations are often called quasilinear because the dispersive term in each equation is linear. The dispersion produced is compensated by nonlinear effects resulting in the formation of exponentially localized solitons. There also exist fully nonlinear evolution equations characterized by nonlinear dispersive terms [24].These fully nonlinear evolution equations support physically interesting solutions which are unattainable with linear dispersion. For example, the equation given by [25]

$$u_t + 3u^2 u_x + 6u_x u_{2x} + 2uu_{3x} = 0 \tag{57}$$

supports solitary wave solutions which are free from characteristic exponential tails of solitons. These solutions are referred to as compactons. Understandably, compactons vanish identically outside a finite but are robust within their range of existence. We shall show that (57) is non-Lagrangian but can be used to define a class of fully nonlinear evolution equations which support compacton solution.

In terms of Casimir potential (57) reads

$$\omega_{xt} = -3\omega_x^2 \omega_{2x} + 6\omega_{2x}\omega_{3x} + 2\omega_x \omega_{4x} \tag{58}$$

for which we can write

$$\omega_t + \omega_x^3 - 2\omega_{2x}^2 - 2\omega_x \omega_{3x} = 0 \tag{59}$$

and
$$P[\omega] = -3\omega_x^2 \omega_{2x} + 6\omega_{2x}\omega_{3x} + 2\omega_x \omega_{4x}. \tag{60}$$

From (60) one can verify that $D_P \neq D_P^*$ such that (57) does not have an analytic representation. However, one can rewrite the Euler-Lagrange expression in the form

$$P[\omega] = \alpha \omega_x^2 \omega_{2x} + \beta \omega_{2x}\omega_{3x} + \gamma \omega_x \omega_{4x} \tag{61}$$

and demand that there exists a choice for the values of $\alpha$, $\beta$ and $\gamma$ such that $D_P$ is self-adjoint.. This viewpoint leads to

$$u_t - \alpha u^2 u_x + \beta u_x u_{2x} + \frac{\beta}{2} u u_{3x} = 0 \tag{62}$$

Note that it is not possible to choose the values of $\alpha$ and $\beta$ to get (57) from (62). Thus (62) provides a new set of nonlinear evolution equations that does not include (57). However it is an interesting curiosity to note that independently of the values of $\alpha$ and $\beta$ supports compecton solutions [18].

## 4. Concluding remarks

Representation of dynamical systems by Lagrangians or the so-called analytic representation plays a role in a wide verities of physical problems ranging from those in classical mechanics to those in quantum field theory. In general, for any given system one can solve the inverse variational problem to construct either the direct or indirect analytic representation. There are, however, systems which admit both representations simultaneously. The most common example in respect of this is provided by the damped harmonic oscillator. In this work we have explicitly demonstrated that self-adjoint form of the equation of motion (discrete system) and/or field equation (continuous system) provides a basis to construct direct analytic representation. But the original evolution equation in conjunction with its adjoint helps us derive the indirect representation. Significantly enough, in point mechanics our approach provided a realization for the Bateman dual system which was introduced by using purely phenomenological arguments with a view to accommodate an open system in the frame of the action principle. With particular attention to nonlinear evolution equations with linear dispersive terms we found that equations in the KdV hierarchy are self adjoint such that each member in the hierarchy can have direct analytic representation. On the other hand nonlinear evolution equations with nonlinear dispersive terms are non-self- adjoint and do not allow one to construct analytic representation. Keeping this in view we started from a typical equation of this type and provided a method to construct a family of nonlinear evolution equations that can be represented by Lagrangians. In this context we note that in refs. 15 and 16 the problem of constructing indirect analytic representation of such equations was solved satisfactorily.